# Optimizing Throughput on Guaranteed-Bandwidth WAN Networks for the Large Synoptic Survey Telescope (LSST)

D. Michael Freemon

National Center for Supercomputing Applications
University of Illinois
Urbana, IL, USA
mfreemon@illinois.edu

*Abstract*—The Large Synoptic Survey Telescope (LSST) is a proposed 8.4-meter telescope that will be located in the Andes mountains in Chile. Every 17 seconds, a 6.4 GB image is transferred to Illinois for immediate processing. That transfer needs to complete within approximately five seconds. LSST is provisioning an international WAN with a 10Gbps bandwidth guarantee for this and other project-related data transfers. The stringent latency requirement drives a re-examination of TCP congestion control for this use case. Specifically, prior work on dedicated Long Fat Networks (LFNs) does not go far enough in fully leveraging the opportunity provided by the bandwidth guarantee. This paper presents an approach for how optimal network throughput can be obtained for the LSST use case, and the conditions under which any project can achieve data throughput rates on long-distance networks approaching wire speed.

## I. Introduction

The Large Synoptic Survey Telescope (LSST) is a proposed large-aperture, wide-field, ground-based telescope that will scan half the sky continuously for 10 years. The 8.4-meter telescope will be located in the Andes mountains in Chile, taking a 6.4 GB image every 17 seconds. Each of those images needs to be transferred to Illinois within approximately five seconds so that processing can be completed in time to get transient alerts notifications distributed to the worldwide astronomical community within 60 seconds. The network design for the project provisions a guaranteed 10 Gbps of bandwidth between La Serena, Chile, and Champaign, IL. With a network latency of 180ms, conventional TCP congestion control would prevent optimal use of the available bandwidth[1].

This paper presents an approach for achieving high bandwidth utilization rates by disabling the kernel-based TCP congestion control and moving the responsibility for allocating the available bandwidth among nodes to an application-level controller. Existing literature discusses some of the fundamental ideas such as disabling TCP slow start and using a fixed congestion window for guaranteed-bandwidth networks. The contribution of this work is to extend those ideas and apply them for the LSST use case by allowing userspace processes to dynamically change the congestion window for each node on a per-route or per-connection basis, and suggesting the introduction of a controller to dynamically manage the bandwidth allocation. Under these conditions, data payload throughput approaches its theoretical limit, and a single LSST image, after compression, can be transferred from South America to North America in under three seconds.

The remainder of this paper is organized as follows. Related work is summarized in Section II. Section III outlines the new congestion control routine. In Section IV, the benchmark testing is described and the results are presented. Section V is a discussion of related topics. In Section VI, future work is suggested. Section VII concludes.

## II. Related Work

The problem of Long Fat Networks (LFN) is described in Martin-Flatin[1], the first three sections of which should be considered prerequisite reading. Of particular note is Section III Case 8, where the authors summarize the characteristics of a 10Gbps link with 120ms RTT.

In Weigle[2], the authors suggest disabling congestion control for traffic on dedicated links. However, the test results reported in that paper are dated and should no longer be considered valid given the technology changes that have occurred since their paper was written.

In Mudambi[3], the authors disable slow start and use a fixed congestion window (cwnd). However, their implementation requires modifications to application-level programs. In addition, their test results do not reflect the value of their proposition. Specifically, their sustained throughput comparison was performed on a LAN-like (13ms RTT) link, not a high-latency link, and other results in their paper include factors unrelated to network performance, such as one of their figures that shows poor results for C-TCP caused by disk I/O constraints.

In Wang[4], the authors keep slow start and allow the system administrator to specify a fixed bandwidth via a statically-assigned kernel module parameter. The bandwidth value is used to calculate a fixed congestion window value. Once the target cwnd is achieved, their algorithm attempts to maintain the cwnd close to the target. However, their test results showed a larger than expected degradation of throughput under 0.01% packet loss.

In Park[5], the authors propose a new congestion control algorithm for fixed bandwidth networks by tracking the historical minimum RTT and using that to calculate the maximum bandwidth, which is then used to set the slow start threshold (ssthresh) to prevent overshoot during slow start. They then hold cwnd at the maximum value until packet loss occurs. With their approach, they keep slow start and cwnd varies too much during congestion avoidance resulting in throughput degradation.

### III. MODIFICATIONS TO CONGESTION CONTROL

The related work does not introduce the notion of application-level control over the management of the guaranteed bandwidth. Wang[4] implements their "Application Based" rate as a kernel module parameter assigned statically at module load time. Mudambi[3] does discuss the distinction between a "control plane" and "data plane", concepts that we leverage here, but they suggest that TCP congestion control be responsible for creating the virtual circuit that it is about to use. We consider that to be one implementation option of a more general design pattern. Importantly, they request virtual circuit rates based only on what the end hosts can support, treating all data equally. This paper introduces the ability for applications to assign different qualities of service to different data flows.

Two essential elements are needed to entertain the approach described in this paper. First, there must be a guarantee of end-to-end bandwidth. Second, all of the hosts connected to the guaranteed-bandwidth network must be under centralized administrative control. If both of these are true, then the transmission rates of sending hosts can be assigned and updated by a controller. This controller is an application-specific component responsible for coordinating any changes to the congestion window on the sending nodes.

TCP congestion control was created to prevent congestive collapse on shared bandwidth networks[6]. It uses packet loss and/or RTT variations as an indicator of network congestion, and slows down the rate of transmission when those situations are encountered. This is the main source of frustration for achieving good performance on fast high-latency networks.

Since TCP congestion control does not apply in the scenario under consideration, we can disable it with no ill effects. Indeed, this is preferable to using novel protocols based on UDP or IP, since we still need the other features provided by TCP, such as reliable delivery, loss detection, and fast retransmit. In addition, no modifications are required of the application-layer software, i.e. the sockets API and semantics remains unchanged. This approach is compatible with all existing file transfer software, include third-party software for which the source code is not available.

The design and implementation of the TCP congestion control module is straightforward. The key features are:
- Use a fixed congestion window (cwnd) for packets being sent to specific address/port ranges. Do not implement slow start and do not decrease cwnd in response to packet loss.
- For packets being sent to all other destinations, use a traditional congestion control algorithm. This enables the same server to concurrently handle traffic to other (shared bandwidth) networks. In our prototype code, we use the algorithm from the Scalable TCP[7] congestion control module.
- Use the /proc filesystem to allow userspace programs to dynamically change the value used for the congestion window (cwnd).

### IV. BENCHMARK TESTING

#### A. Equipment Setup

The equipment setup is depicted in Figure 1. All machines are Fedora 17. The sending machines, with the modified congestion control, are running linux kernel 3.7.9. The hardware used is unremarkable: the servers have a single dual core Intel 6600 CPU running at 2.4GHz and 2 GB of DDR2 memory. All network interface cards are 1Gbps with an MTU of 1500 bytes. The network switch is an entry-level consumer-grade Trendnet TEG-S50g.

Netem[8] is used to simulate the wide-area network between La Serena, Chile, and Champaign, IL. For latency, the National Optical Astronomy Observatory (NOAO) has historical information for the existing networks between North America and La Serena that shows an average of 180ms, a value that we adopt for this testing. For packet loss, typical service level agreements with the South American network providers establish a maximum threshold at 0.01% in each direction, above which the providers consider the network to be unavailable, triggering financial penalties. We adopt 0.01% in this paper, which represents the worst-case likely to be seen during normal operations. The specific netem settings are:

$ tc qdisc show
qdisc netem 8002: dev p3p1 root refcnt 2 limit 50000 delay 90.0ms loss 0.01%
qdisc netem 8001: dev p5p1 root refcnt 2 limit 50000 delay 90.0ms loss 0.01%

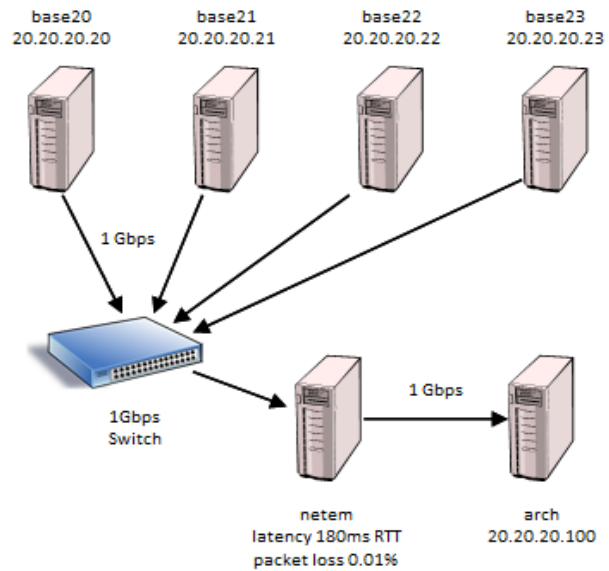

Figure 1. The equipment setup for the benchmark testing.

All sending nodes (base20-base23) have the modified congestion control routine installed. In addition to the normal system tuning of TCP/IP, kernel parameters tcp_frto and tcp_frto_response were both set to 2. Additionally, the initial receive window on the receiving system was set to 99999 using the ip route command. The nuttcp program[9], version 7.1.6, was used to generate all network traffic.

## B. The LSST Scenario

LSST has three different types of flows: Urgent file-based data which must be transferred immediately and quickly (the crosstalk-corrected images), file-based data which can be transferred as low-priority background traffic (raw images and all other file-based data), and unmodified TCP/IP socket applications that must share the same network and receive predictable performance undisturbed by other traffic. MySQL replication of the Engineering and Facility Database (EFD) is an example of the third type.

The first two machines, base20 and base21, are assigned to transmit crosstalk-corrected images. Base22 is the MySQL database server that is continuously replicating to the Arch server. Base23 is responsible for sending raw images in the background.

Bandwidth management is performed by dynamically manipulating the congestion window (cwnd) of the sending servers. Testing showed that the optimal total cwnd for this network configuration is 14,764 packets. Throughout the test, the MySQL/EFD replication (base22) is given a cwnd of 738, or 5% of the bandwidth. As the test begins, there is no crosstalk-corrected images to transfer, so the background transfer (base23) is given the remaining 95% of the bandwidth (cwnd 14026). At some point, it becomes known to the controller that a crosstalk-corrected image needs to be transferred. In preparation for that transfer, the cwnd is reduced on the background transfer (base23) to 148 (or 1% of the bandwidth). Then the crosstalk-corrected image transfers begin. Base20 and base21 both send one-half of the image with their cwnd of 6939 (or 47% each). After the crosstalk-corrected image transfers are complete, the bandwidth assigned to the background transfer (base 23) is brought back up to the 95% level.

## C. Test Results

Figure 2 shows the test results. The gap at Second 5 and again at Second 30 is larger than necessary due to overheads that would not be present in a production system. A separate remote server was used to run a shell script to serve as a surrogate software controller, which uses SSH to remotely change cwnd values and launch nuttcp processes. Much of the time in those gaps was spent in connection establishment and authentication, which is time that can be significantly reduced with better control software.

The vertical axis is the payload data being received by the destination server. The maximum possible payload throughput over a 1Gbps network with a MSS of 1448 is 117.6 MB/s (941 Mbps). Looking at the "all" line, which shows the aggregate of all incoming traffic, we see payload data rates of 117.1 MB/s (937 Mbps), or 99.6% of the theoretical maximum. This represents a network utilization (the ratio of payload speed to wire speed) of 94%.

For the crosstalk-corrected file transfers, the following are the commands that were executed along with their corresponding output:

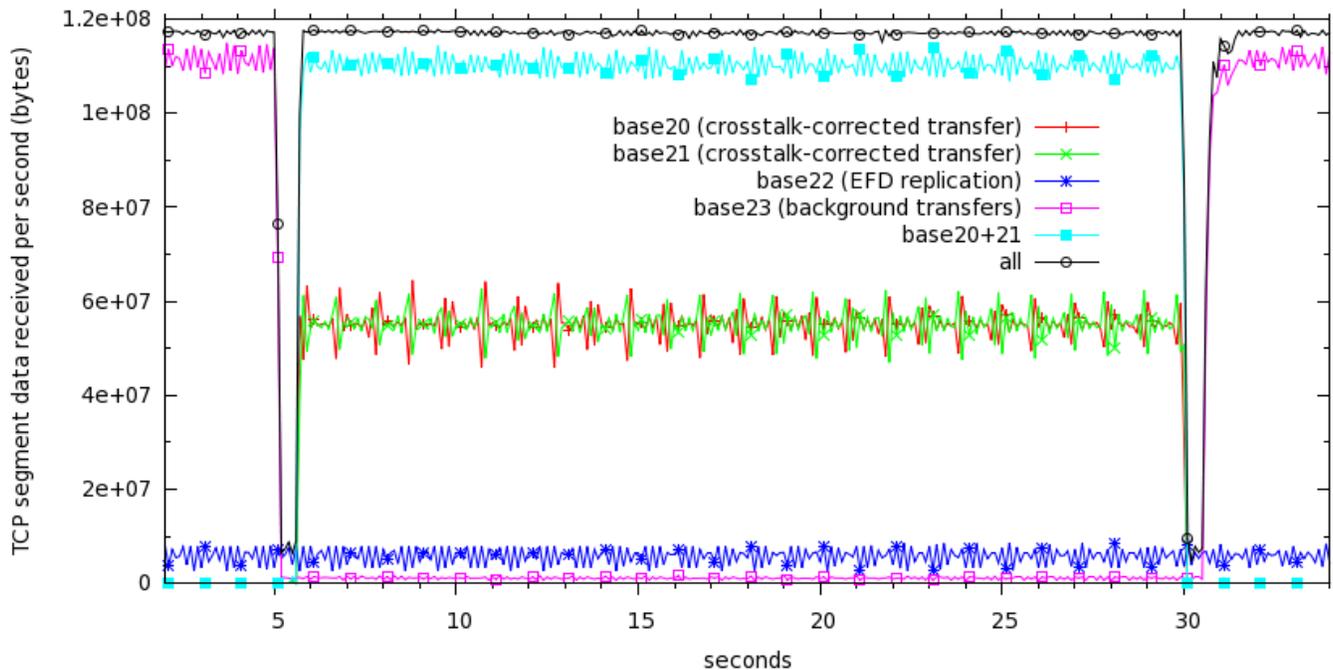

Figure 2. Payload throughput, in bytes per second, as measured from the receiving system, over a 1Gbps network with 180ms of latency (RTT) and 0.01% packet loss (each way).

```
~/nuttcp-7.1.6 -Ibase20 –w128m -n 20508 -p 5101 20.20.20.100
~/nuttcp-7.1.6 -Ibase21 –w128m -n 20508 -p 5201 20.20.20.100
base20: 1281.7500 MB /  24.70 sec =  435.3043 Mbps 1 %TX 6 %RX 104 retrans 180.27 msRTT
base21: 1281.7500 MB /  24.71 sec =  435.0694 Mbps 2 %TX 6 %RX 126 retrans 180.33 msRTT
```

The nuttcp output indicates 2563.5 MB (1281.75 x 2), which is MebiBytes. As confirmation, Wireshark shows, via the TCP ACK numbers, that 1,344,012,290 bytes were transferred from each of base20 and base21, for a total of 2688MB (decimal). This is equivalent to a single 6.4 GB LSST image after compression. The compression ratio has previously been validated as correct for this type of file by the LSST Data Management team.

The crosstalk-corrected transfers completed in 24.5 seconds (see discussion below for why this is different than that reported by nuttcp). For a 10Gbps network, the transfer time would be 2.45 seconds.

It should be noted that there were retransmissions as a result of packet loss. The nuttcp output and the TCP ACKs account for that and report only "goodput". The retransmission count reported by nuttcp for the crosstalk-corrected transfers is consistent with what we expect. There were 2.046 million packets sent, of which 230 (or ~0.01%) were dropped in transit and retransmitted. Note that 0.01% of the ACK packets being returned to the sender were also dropped, but those do not necessarily result in a retransmission of data.

Throughout the whole test, the EFD (base22) is replicating at approximately 5.8 MB/s. It is also worth noting explicitly that these changes are dynamic, even for existing TCP sessions. Base23 in Figure 2 is a continuously-open connection sending data.

## V. DISCUSSION

### A. Initial Burst

When the coordinated set of servers begins sending the crosstalk-corrected data at exactly the same time, there is an initial burst of packets than can overwhelm the network switch that multiplexes the traffic. There are a number of ways to avoid an excessive number of dropped packets at that switch. The first, which was used during the testing in this paper, is to use Ethernet PAUSE frames. The head-of-line blocking that results is of no concern to us in this scenario. Our testing has shown that PAUSE frames do not introduce any particular problems, nor do they cause any noticeable slowdown in throughput.

There are a number of additional techniques that can be employed should we seek to minimize the use of Ethernet PAUSE frames. They include switch buffering, NICs speeds, and sender-side rate throttling. If only two servers are transmitting, then 11MB of switch buffer would be sufficient to avoid PAUSE frames. With 10 servers transmitting, 20MB of switch buffer is needed, with 23MB of buffer being the worst case in any configuration. Even today (2013), switches are available with that much buffer space. Alternatively, we could simply use 1Gbps NICs on the 10 sending servers. Or if the sending servers already have 10Gbps NICs installed, the speed of those interface cards could be set to 1Gbps. And finally, we could configure rate throttling in the sending servers. We tested our setup using the Hierarchy Token Bucket packet scheduler[15] to slow down the sending rate, which we found to work well. Since the results were identical to the use of PAUSE frames, we discontinued our use of HTB and did not use it for any of the test results reported in this paper.

### B. Extrapolating to 10Gbps

The testing described in this paper uses 1Gbps hardware and extrapolates to the 10Gbps scenario by dividing by 10. We conclude that this is a valid assumption for the reasons described in this section. Although 10Gbps networking will be commonplace by the time LSST is scheduled to go into Operations, it is not necessary to show that a single host can send or receive traffic at the full rate of 10Gbps. LSST will have multiple servers sending and receiving only fractional portions of the full network traffic. This immediately disposes of concerns regarding system resource constraints associated with a single host, such as filesystem performance, interrupt processing/coalescence, etc.

A potential concern is that latency does not scale in proportion to bandwidth. Although a 10Gbps network is faster, the latency remains the same. We address this concern with the following test: Consider 10 hosts each sending 10% of the crosstalk-corrected image on 10% of the bandwidth of a 10Gbps network. Figure 3 shows the results of one such slice. The transfer of 268.8MB of data, with 180ms of latency (RTT) and 0.01% packet loss (each way), is completed in 2.65 seconds. The small spike and pause for a duration of one RTT at the 1 second mark on Figure 3 is caused by the TCP window scaling option not being in effect for the receive window until receipt of the first ACK after the TCP handshake[14]. This initial pause is easily avoided by reusing TCP connections, in which case the time to transfer this data would have been 2.47 seconds. Figure 3 shows only data movement, and not the extra RTT required to confirm the transfer, which is included in the 2.47 and also reported by nuttcp. The nuttcp command and output is:

```
$ ./nuttcp-7.1.6 -w512m -n 4102 20.20.20.100
   256.3750 MB /  2.65 sec =  812.1586 Mbps 100 %TX
13 %RX 22 retrans 180.27 msRTT
```

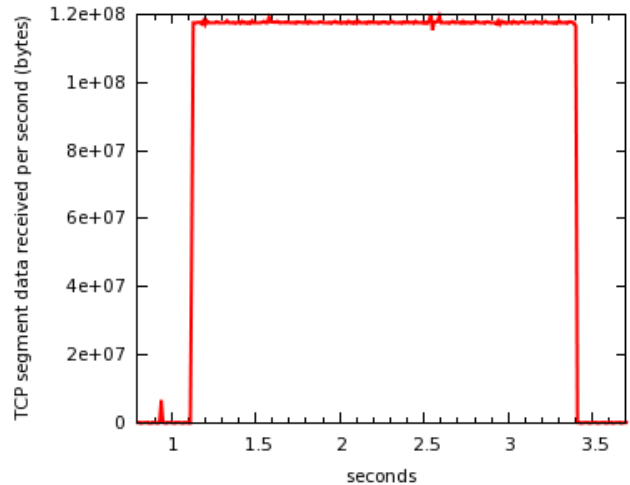

Figure 3. Transferring 10% of the compressed crosstalk-corrected image on a 1Gbps network with 180ms latency at 0.01% packet loss.

*C. Sensitivity to Variations in Latency and Packet Loss*

We have assumed throughout this paper a latency value of 180ms and packet loss rate of 0.01% in each direction, and have argued that the former is a realistic estimate based on real-world measurements, and the latter is a worst-case scenario. To show the sensitivity of the benchmark results to variations in those assumptions, the test in the preceding section was performed for a range of latencies and packet loss percentages. Table 1 provides the results as reported by nuttcp.

As discussed earlier in this paper, the time needed by TCP to enable window scaling (1 RTT) is easily avoided. Table 2 and Figure 4 show the median times from Table 1 with the additional assumption that TCP connections are reused or otherwise initialized such that the receive window is set as needed before data transmission is initiated.

*D. Retransmission Tail*

In all of the tests conducted for Table 1, the throughput pattern is the same as that shown in Figure 3. If we imagine the shape of the data in Figure 3 as a box, then the width of the box is exactly 2.30 seconds or less for all the runs listed in Table 1. Packet loss rates influence the width of the box only very slightly, and latency does not affect it at all. Even at 10% packet loss (not shown), the width of the box is only 2.66 seconds.

The higher transfer times at the higher latency and loss rates (the bottom right portions of Table 1 and Table 2) are caused by a "retransmission tail". This is a result of retransmission timeouts, with latency delays aggravated by the binary exponential backoff of the retransmission timer. For example, in one of the runs with a RTT of 800ms and a packet loss rate of 1%, the box width was 2.30 seconds, as expected, but it took an additional 4.42 seconds to get the final bytes transferred and acknowledged. We believe this retransmission tail can be reduced to just a just a few RTT times. Implementing and testing this is left as future work.

*E. Parallel TCP Streams*

Most file transfer solutions (e.g. FDT[10], GridFTP[11], bbcp[16]) use multiple TCP streams to minimize the negative impact of packet loss on throughput caused by traditional congestion control algorithms. With the approach described in this paper, it is not necessary to use parallel TCP streams from the same host to achieve optimal network throughput. There remains benefits to using multiple hosts in parallel, some of which have been previously mentioned, but not multiple streams per host. However, applications are not precluded from using multiple streams per host if their software design suggests it for other reasons.

*F. Jumbo Frames*

This paper uses a MTU of 1500 bytes, which results in 6% protocol overhead. From the vantage point of the year 2013, it is difficult to imagine that we will not have the option of using 9000 byte MTUs by the year 2020, when LSST is scheduled to go into Commissioning and Operations. The use of such jumbo frames will reduce protocol overhead to 1%, thereby increasing data transfer throughput rates described in this paper by 5%.

| Latency (ms) | Packet Loss 0% | | | 0.001% | | | 0.01% | | | 0.1% | | | 1% | | |
|---|---|---|---|---|---|---|---|---|---|---|---|---|---|---|---|
| 0 | 2.29 | 2.29 | 0.00 | 2.29 | 2.29 | 0.00 | 2.29 | 2.29 | 0.00 | 2.29 | 2.29 | 0.00 | 2.31 | 2.33 | 0.06 |
| 50 | 2.39 | 2.39 | 0.00 | 2.39 | 2.39 | 0.00 | 2.39 | 2.39 | 0.01 | 2.39 | 2.40 | 0.01 | 2.46 | 2.57 | 0.19 |
| 100 | 2.49 | 2.49 | 0.00 | 2.49 | 2.49 | 0.00 | 2.49 | 2.49 | 0.02 | 2.53 | 2.53 | 0.04 | 2.61 | 2.69 | 0.18 |
| 150 | 2.59 | 2.59 | 0.00 | 2.59 | 2.60 | 0.03 | 2.59 | 2.61 | 0.04 | 2.69 | 2.70 | 0.15 | 2.75 | 2.89 | 0.30 |
| 180 | 2.65 | 2.65 | 0.00 | 2.65 | 2.67 | 0.08 | 2.65 | 2.67 | 0.05 | 2.75 | 2.84 | 0.43 | 2.85 | 3.04 | 0.40 |
| 200 | 2.69 | 2.69 | 0.00 | 2.69 | 2.69 | 0.00 | 2.69 | 2.72 | 0.06 | 2.80 | 2.88 | 0.33 | 2.91 | 3.11 | 0.33 |
| 300 | 2.89 | 2.89 | 0.00 | 2.89 | 2.89 | 0.00 | 2.89 | 2.91 | 0.06 | 3.10 | 3.06 | 0.11 | 3.21 | 3.59 | 0.55 |
| 400 | 3.09 | 3.09 | 0.00 | 3.09 | 3.10 | 0.05 | 3.09 | 3.15 | 0.11 | 3.42 | 3.48 | 0.41 | 3.51 | 4.16 | 0.93 |
| 500 | 3.29 | 3.29 | 0.00 | 3.29 | 3.31 | 0.07 | 3.29 | 3.34 | 0.12 | 3.72 | 3.82 | 0.53 | 5.49 | 5.17 | 0.98 |
| 600 | 3.49 | 3.49 | 0.00 | 3.49 | 3.51 | 0.11 | 3.49 | 3.67 | 0.30 | 4.03 | 3.98 | 0.10 | 5.97 | 5.56 | 1.30 |
| 700 | 3.69 | 3.69 | 0.00 | 3.69 | 3.70 | 0.05 | 3.69 | 3.84 | 0.21 | 4.26 | 4.26 | 0.08 | 6.93 | 6.04 | 1.51 |
| 800 | 3.89 | 3.89 | 0.00 | 3.89 | 3.92 | 0.15 | 3.89 | 4.04 | 0.24 | 4.60 | 4.57 | 0.12 | 7.95 | 7.57 | 1.60 |

Table 1. Mean, median, and standard deviation time, in seconds, to transfer 10% of a compressed crosstalk-corrected image on a 1Gbps network for the indicated latencies and packet loss rates. The results of 20 consecutive runs were included in each entry. These results include the extra time needed by TCP to adjust the receive window for new connections to use the window scaling factor.

*G. Application Control*

This paper is focused on optimizing the network throughput in the context of a specific application. This is necessary as the approach described here is not a general service whose run-time behavior can be implemented independently of the application using it. The application controller or system administrator must be aware of all processes allowed to use a fixed congestion window. Each application system needs to assess its ability to meet this requirement as part of the application system design process.

*H. Latency Variability*

In this paper, the congestion window was calculated based on an unchanging latency (RTT) of 180ms. On real networks, many factors cause the latency to change from the expected baseline value. With respect to the application's requirements, the crosstalk-corrected transfers do not miss their required time deadline until latency exceeds 360ms. At higher latencies, the impact on payload throughput is inversely proportional (twice the latency results in a halving of payload throughput), so one could argue that a fixed latency assumption for this specific application use case is a good enough. There are no requirements problems in the case where latency is lower than the expected value. However, in practice, we would chose to improve how the latency variations are handled, which is included in the future work section.

*I. Advertised Bandwidth Guarantees by Network Providers*

The bandwidth guarantees mentioned throughout this paper is the guarantee as assumed by the application system. This does not need to match exactly with the bandwidth guarantee advertised by the network provider. There are a number of technical and policy issues that can prevent a network from delivering what it has guaranteed. Despite our reluctance to simply accept such a situation, we can accommodate it by using a slightly lower bandwidth assumption in the application controller. For example, if the network vendor guarantees 10Gbps, the application system could assume the guarantee is 9.9Gbps.

*J. Compression Time*

The LSST raw image data is stored in per-amp files of approximately 2MB each. Using FPACK[13] with the compression options planned by LSST, tests show that it takes 0.02 seconds to compress it. Since the files will be compressed in parallel, the whole process will take about 0.02 seconds. As a worst case, if per-CCD files were used, which would be approximately 32MB in size, the compression time would be in the range of 0.3 to 0.4 seconds. In both cases, we can accommodate it in the current time budget requirements for the transfer.

## VI. FUTURE WORK

There are a number of operational and usability improvements that can be made to the congestion control code used in this testing. The existing code serves only as a proof-of-concept. Specific areas of improvement include adding flexibility to how the connections that are subject to the fixed congestion window are specified, and allowing the system administrator to specify an existing congestion control module for all other connections (changing this module into a "stacked" kernel module). Adjusting the congestion window for changes in latency (RTT) would likely be performed in the congestion control code, although other designs exist that would make this adjustment elsewhere. There should also be some additional code to guard against extreme conditions, such as falling back to normal congestion control if packet loss begins to occur at catastrophic rates.

| Latency (ms) | Packet Loss | | | | |
|---|---|---|---|---|---|
| | 0% | 0.001% | 0.01% | 0.1% | 1% |
| 0 | 2.29 | 2.29 | 2.29 | 2.29 | 2.31 |
| 50 | 2.34 | 2.34 | 2.34 | 2.34 | 2.41 |
| 100 | 2.39 | 2.39 | 2.39 | 2.43 | 2.51 |
| 150 | 2.44 | 2.44 | 2.44 | 2.54 | 2.60 |
| 180 | 2.47 | 2.47 | 2.47 | 2.57 | 2.67 |
| 200 | 2.49 | 2.49 | 2.49 | 2.60 | 2.71 |
| 300 | 2.59 | 2.59 | 2.59 | 2.80 | 2.91 |
| 400 | 2.69 | 2.69 | 2.69 | 3.02 | 3.11 |
| 500 | 2.79 | 2.79 | 2.79 | 3.22 | 4.99 |
| 600 | 2.89 | 2.89 | 2.89 | 3.43 | 5.37 |
| 700 | 2.99 | 2.99 | 2.99 | 3.56 | 6.23 |
| 800 | 3.09 | 3.09 | 3.09 | 3.80 | 7.15 |

Table 2. Median time, in seconds, to transfer 10% of a compressed crosstalk-corrected image on a 1Gbps network for the indicated latencies and packet loss rates.

This paper does not discuss the controller design in any detail, but that is an important component of any application system leveraging this approach. Such a controller could, optionally, work in concert with a dynamic network provisioning system such as OSCARS[12] to ensure the required bandwidth guarantee. In the special case where there is a one-to-one relationship between TCP sessions and virtual circuits, then the cwnd can be set at a maximum, effectively transferring control of bandwidth management from the congestion control routine to the network devices. In the alternative, some projects may elect to design their system such that congestion window assignments do not need to be changed dynamically, thus eliminating the need for a controller.

As with all testing conducted by simulating certain network characteristics, it is valuable to conduct the same benchmarks tests on real long-distance networks whenever possible. For example, packet loss in this paper was assumed to be random, whereas testing on the real network is likely to reveal correlations in the distribution of packet loss. Additionally, should LSST encounter packet loss rates far in excess of what is currently expected, then future work should include RTO modifications to reduce the retransmission tail, which provides lower and more predictable transfer times at those very high packet loss rates.

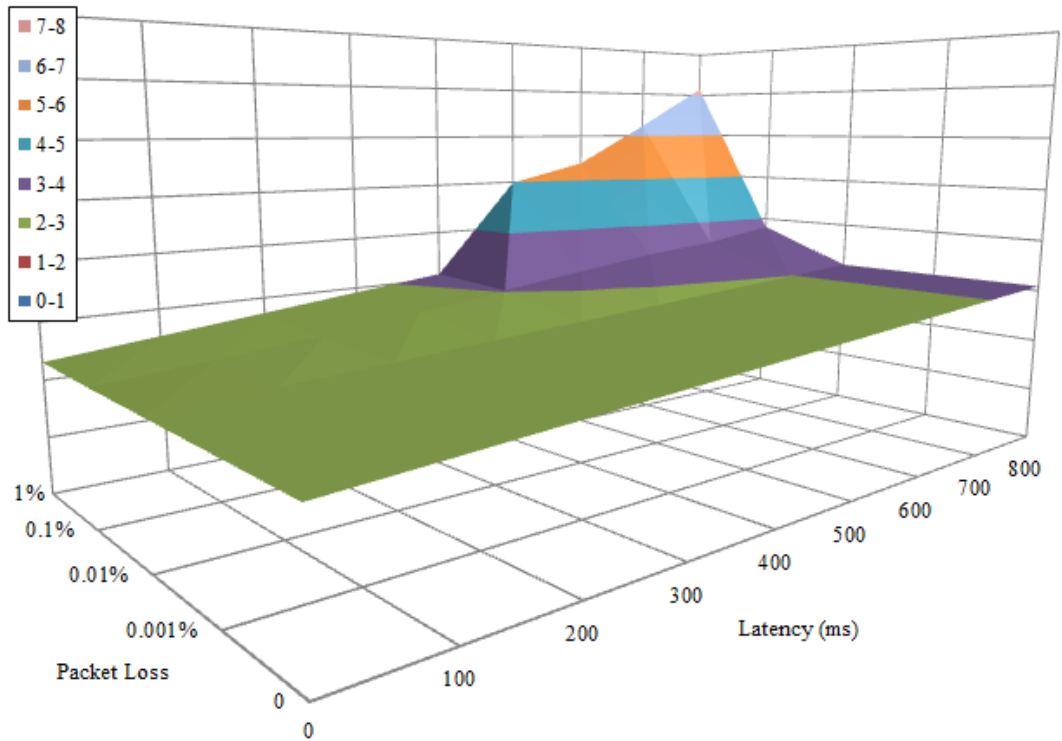

Figure 4. Median time, in seconds, to transfer 10% of a compressed crosstalk-corrected image on a 1Gbps network for the indicated latencies and packet loss rates.

## VII. CONCLUSION

In the era of big data, extracting the most value from expensive long-haul networks is critical. This paper has presented an approach where, if bandwidth is guaranteed and connected hosts are managed, then data can be transferred at full link speed over high-latency international networks.

This approach to bandwidth management is transparent to userspace processes, and is compatible with all existing file transfer applications and utilities. No application-layer code needs to be modified or recompiled.

The LSST project provided the use case scenarios discussed in this paper, but the solution can be adopted by any project or system if the necessary prerequisites are met.

## ACKNOWLEDGMENT

The author would like to thank those who provided valuable feedback on early drafts of this paper, including Gregory Dubois-Felsmann, Jeff Kantor, Kian-Tat Lim, and Darrell Newcomb.